\documentclass[english,showpacs,twocolumn,amsmath,amssymb,longbibliography]{revtex4-1}
\usepackage{lineno}

\usepackage[english]{babel}
\usepackage{xcolor}
\usepackage{epsfig}
\usepackage{float}
\usepackage{graphicx}
\usepackage{bm}
\usepackage{array}

\begin{document}

\title{Random networks with $q$-exponential degree distribution}

\author{Cesar I. N. Sampaio Filho, Marcio M. Bastos, Hans J. Herrmann, Andr\'e A. Moreira, and  Jos\'e S. Andrade Jr.}
\affiliation{Departamento de F\'isica, Universidade Federal do
  Cear\'a, 60451-970 Fortaleza, Brazil}

\begin{abstract}
We use the configuration model to generate networks having a degree distribution 
that follows a $q$-exponential, $P_q(k)=(2-q)\lambda[1-(1-q)\lambda k]^{1/(q-1)}$ , for arbitrary values of the parameters $q$ and $\lambda$.
We study the assortativity and the shortest path of these networks finding that the more
the distribution resembles a pure power law, the less well connected are the corresponding nodes. In fact, the average degree of a nearest neighbor grows monotonically with $\lambda^{-1}$. Moreover, our results show that $q$-exponential networks are more robust against random failures and against malicious attacks than standard scale-free networks. Indeed, the critical fraction of removed  nodes grows logarithmically with $\lambda^{-1}$ for malicious attacks. An analysis of the  $k_s$-core decomposition shows that $q$-exponential networks have a highest $k_s$-core, that is bigger and has a larger $k_s$ than pure scale-free networks. Being at the same time well connected and robust, networks with $q$-exponential degree distribution exhibit scale-free and small-world properties, making them a particularly suitable model for application in several systems.

\end{abstract}\maketitle

In a large variety of fields one finds experiments, numerical results and theoretical models 
that fairly well agree with $q$-exponentials. This includes applications in fully developed
turbulence~\cite{Beck1}, anomalous diffusion in plasmas~\cite{Liu}, statistics of cosmic
rays~\cite{Beck2}, econometry~\cite{Souza, Borges}, biophysics~\cite{Upadhyaya} and many others.
In particular, many empirical complex networks have been found to follow $q$-exponential degree
distributions~\cite{Carro, Thurner1, Wedemann}. The same behavior has also been detected in
several proposed model networks~\cite{Thurner2}, most of them generated through growth models based on the
preferential attachment  principle~\cite{Soares, Brito1, Brito2, Ochiai}. These models produce
empirically $q$-exponential distributions, which in some cases can even be confirmed
analytically~\cite{Thurner1, Ochiai}. But what is still missing is a systematic way of producing random networks with
arbitrarily chosen $q$-exponential degree distributions. Here will present such a method based on
the configuration model~\cite{Newmanbook} and study the properties of the
resulting networks. 

In a network the degree $k$ of a node is defined as the number of connections it has with other
nodes. Networks are called scale-free when the distribution of degrees follows a power law with
exponent $\gamma$, $P(k) \sim k^{-\gamma}$. The $q$-exponential distribution  given by,
\begin{equation}\label{eq01}
P_q(k)=(2-q)\lambda[1-(1-q)\lambda k]^{1/(q-1)},
\end{equation}
has two parameters, $q \geq 1$ and $\lambda \geq 0$. It is a generalization of a power-law distribution, since for large $k$ ($k\gg\lambda^{-1}$) it decays like $k^{-\gamma}$ with
$\gamma = 1/(q-1)$. For small degrees ($k\ll \lambda^{-1}$), however, it tends to a plateau distribution of height $2-q$. The parameter $\lambda^{-1}$ therefore determines the crossover between these two regimes. The distribution Eq.~\ref{eq01} represents a fundamental ingredient in the mathematical formalism of the generalized thermostatistics and its applications~\cite{tsallis2009introduction,adib2003tsallis,andrade2002extended,hasegawa2009bose,hasegawa2010interpolation,andrade2010thermostatistics,nobre2011nonlinear,
hanel2011comprehensive,nobre2012generalized}. 

Although most properties observed in complex networks with heavy-tailed distributions
are determined by the shape of the tail, deviations from the power law at smaller degrees 
alter the occurrence of the least connected nodes, resulting in structural changes 
that can affect processes taking place on these networks. In what follows, we take a closer look at these effects.

Different from previous works  based on preferential attachment models of growth to obtain networks with degree distributions mimicking
the $q$-exponential behavior, here, as already mentioned, we employ the configuration model~\cite{Newmanbook} to build our random networks.  
In this way, we assure that the particular topological properties observed, including intrinsic node correlations, are not induced by the growth process,
but are rather a direct consequence of the $q$-exponential form of the degree distribution.

\begin{figure}[t]
\includegraphics*[width=\columnwidth]{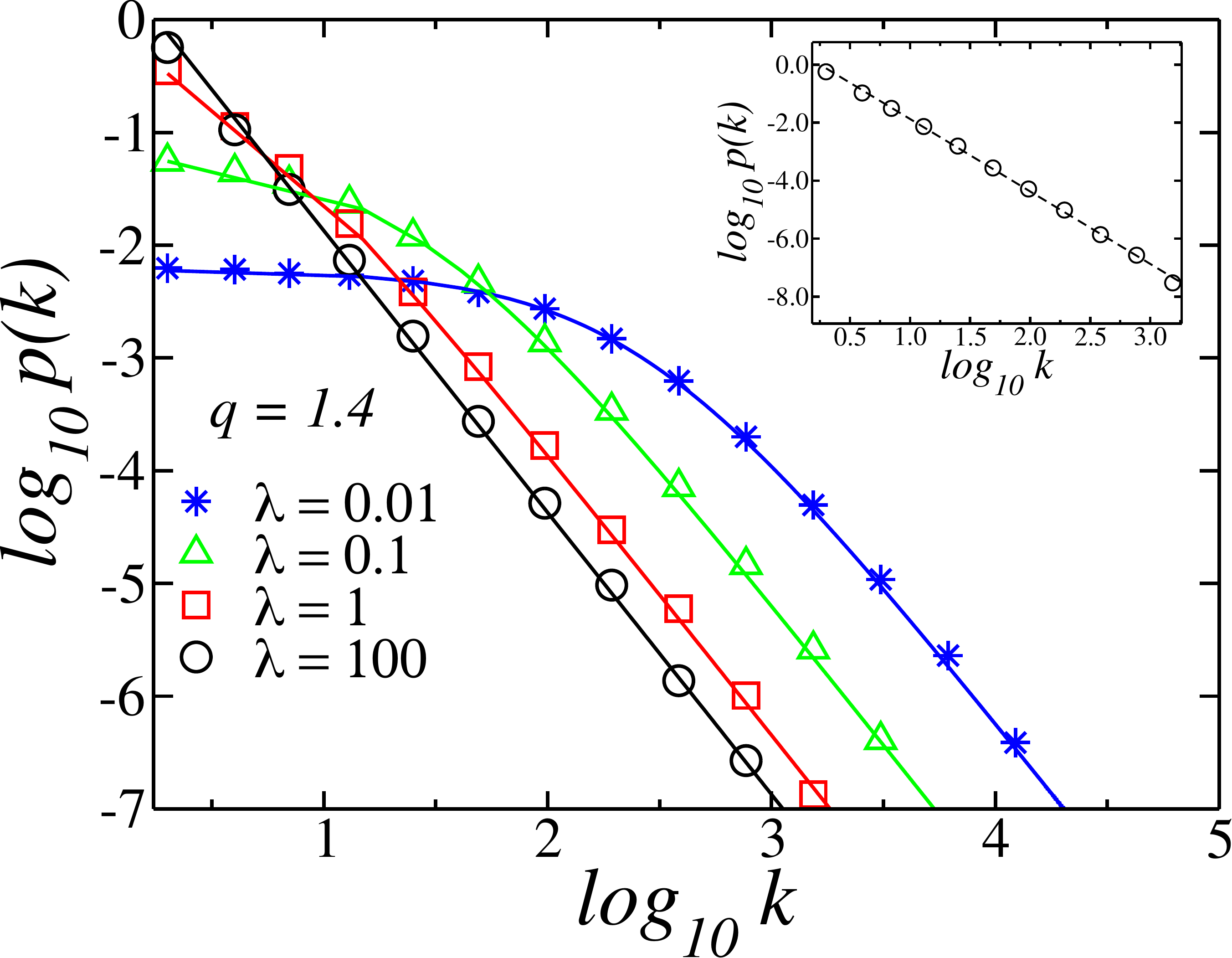}
\caption{Degree distribution of $q$-exponential networks (symbols) compared with the
expected distribution of Eq.~\ref{eq01} (solid lines) for $q = 1.4$ and 
for $\lambda=0.01$ (blue stars), $\lambda=0.1$ (green triangles), $\lambda=1$ 
(red squares), and $\lambda=100$ (black circles). The inset shows a comparison between
the degree distributions of a $q$-exponential with $\lambda=100$ and a pure scale-free
distribution (dashed black line) with $q=1.4$ ($\gamma=2.5$). The results are obtained for networks
with size $N = 500000$ by averaging over $100$ samples. As can be seen, for small $\lambda$ the distributions
attain a plateau at small degrees and the power-law regime becomes larger with increasing $\lambda$.}
\label{fig01}
\end{figure}

We start by assigning to each node $i$ a prescribed degree $k_i$, drawn from a
$q$-exponential distribution. More precisely, since the degrees are
integers, we chose randomly a number $x_i$ from a $q$-exponential
distribution and define $k_i$ as the largest
integer smaller than $x_i$. To avoid obtaining many small disconnected clusters,
we only consider nodes with degree $k_i \ge 2$, {\it i.e.}, $k_{min} = 2$. We visualize the yet
unconnected $k_i$ degrees on node $i$ by $k_i$ ``stubs". Then we proceed to connect nodes
pairwise. To do this, we choose two different nodes with probabilities
proportional to their number of stubs. If these two nodes are not yet
connected, a link is placed between them, and the number of
stubs if decreased for each of the two nodes by one. This process continues until all stubs
have been connected, or, in case that at the end some stubs remain
that could not be connected, these stubs are removed and the degree originally assigned 
to the corresponding nodes is adjusted accordingly. Furthermore, a maximum degree $k_{max}$ exists
due to the fact that the number of nodes in the network is finite. When $\gamma<3$, the second moment
$\langle k^2\rangle$ of the distribution diverges and one can show that $\langle k^2\rangle\sim k_{max}^{3-\gamma}$, {\it i.e.},  that the
second moment is controlled by the most connected node. 
This property can have profound effects on dynamical
processes taking place on the network, including fragility to targeted attacks and resilience 
to random failure~\cite{Cohen,Moreira,Schneider}.

As we show in Fig.~\ref{fig01}, the degree
distributions obtained by this method follow very closely the expected $q$-exponential form
of Eq.(1). For $\lambda\ll 1$ one observes a pronounced plateau for small $k$. On the other hand,
when $\lambda\gg 1$, effectively, our degree distribution will be identical to a scale-free
distribution with the same $k_{min}$. Therefore, by decreasing the
parameter $\lambda$, we can continuously move away from a scale-free degree distribution 
and widen the plateau. Thus varying $\lambda$ will allow us to identify the
effect of the deviations from pure scale-freeness that are particular to $q$-exponential
distributions.

As already mentioned, the difference between the $q$-exponential and the simple (power-law) scale-free
distribution, is that at a given scale the $q$-exponential crosses-over to a plateau of constant
height. Therefore, on one hand scale-free networks have many more least connected nodes than networks
having $q$-exponential degree distribution with small $\lambda$. 
The smaller $\lambda$, the denser the networks become, increasing the number and degree of their hubs, 
which as we will see next, makes the networks more robust. Topological differences like these can lead
to substantial changes in structural properties of complex networks as well as in the static and dynamical
behavior of models when implemented on these substrates. Next, we
investigate targeted attacks and random failure in $q$-exponential networks. We also measure the intrinsic assortativity of
these networks and their average minimum path. Finally, we employ the $k$-core method to determine the
different structural configurations of the $q$-exponential networks.

\begin{figure}[t]
\includegraphics*[width=\columnwidth]{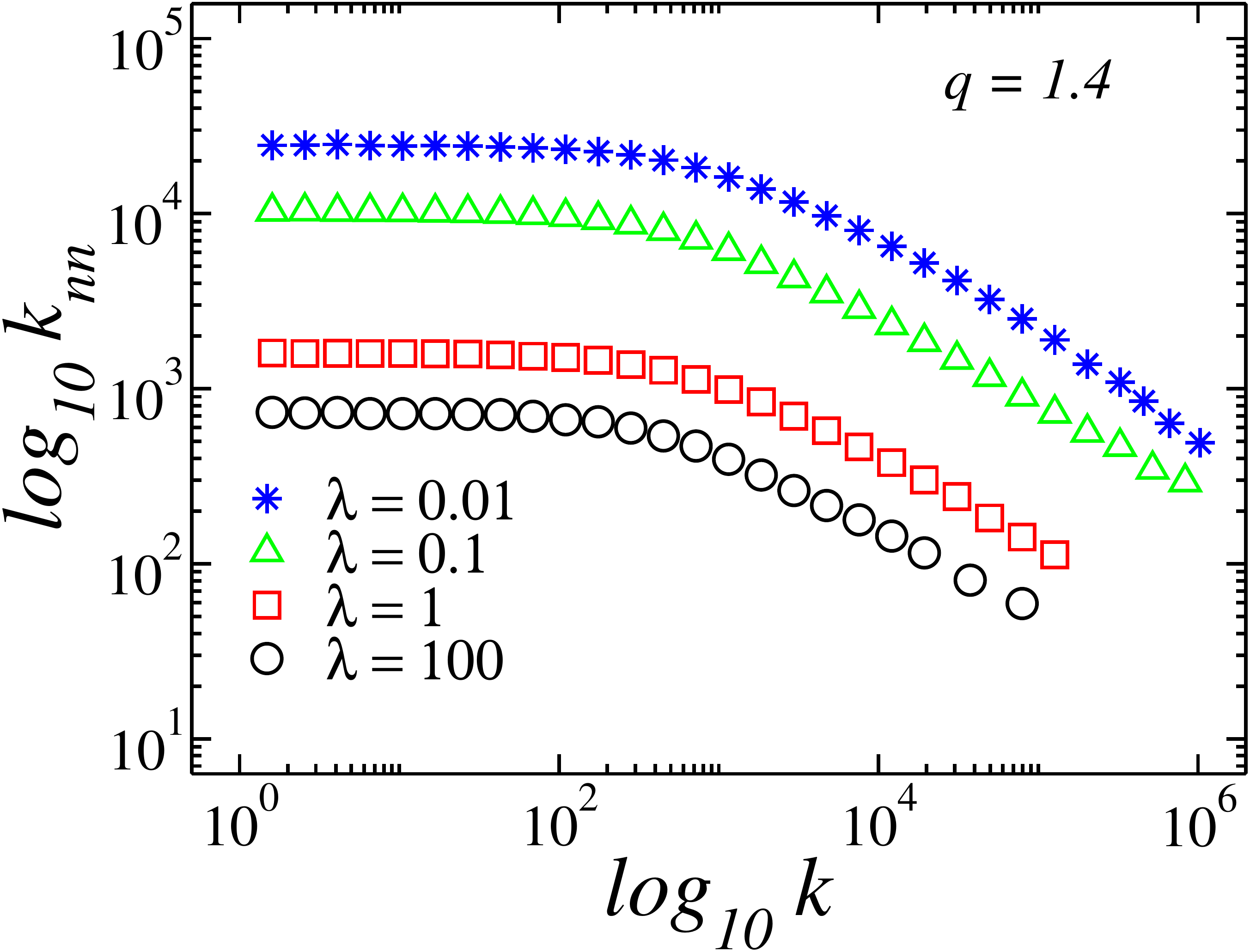}
\caption{Dependence of the average nearest-neighbor degree $k_{nn}$ of vertices on the degree $k$ for networks with size $N=10^{7}$ nodes, $q=1.4$,
and $\lambda = 100$ (black circle), $\lambda = 1$ (red square), $\lambda=0.1$ (green triangle), and $\lambda=0.01$ (blue star).
The baseline of each curve corresponds to the value $\left\langle k^{2} \right\rangle$/$\left\langle k \right\rangle$. All curves point towards intrinsic dissortative behavior for sufficiently large values of $k$.}
\label{fig02}
\end{figure}
\begin{figure}[t]
  \includegraphics[width=\columnwidth]{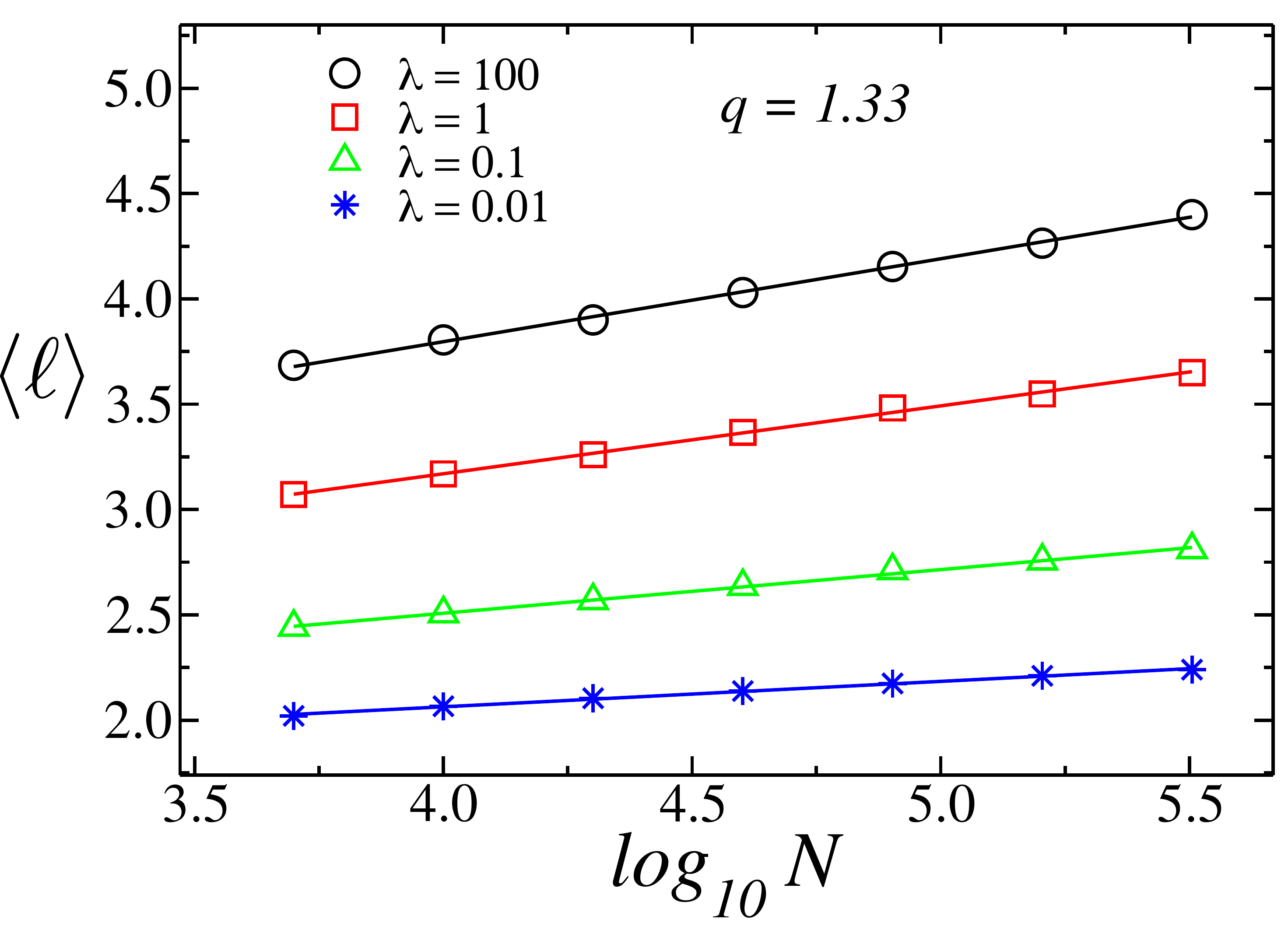}
  \caption{Size dependence of the mean length of the shortest-path $\left\langle \ell \right\rangle$ for $q = 1.33$ and for
$\lambda=0.01$ (blue stars) , $\lambda=0.1$ (green triangles), $\lambda=1$ (red squares), and $\lambda=100$ (black circles).  The
results are obtained by averaging over $10^{5}, 10^{4},10^{4},10^{4},10^{4}$, $10^{3}$, and $10^{3}$ samples of size $N=5000, 10000, 20000, 40000, 80000, 160000,$ and $320000$, respectively.}
\label{fig03}
\end{figure}

It has been shown \cite{newman2003mixing,newman2002assortative,catanzaro2005generation} that random
networks with heavy-tailed degree distribution often display intrinsic assortativity. This is also
the case for $q$-exponential networks. In Fig.~\ref{fig02}, we show the expected average degree
$k_{nn}(k)$ of the nearest neighbors of a given node of degree $k$. As can be seen, the most connected
nodes have a smaller $k_{nn}$, pointing towards negative intrinsic assortativity. The reason for this
dissortative behavior in random networks is that the most connected nodes can neither connect to
themselves nor have multiple connections between them. We also see from Fig.~\ref{fig02} that
$k_{nn}$ is the larger the more the distribution deviates from a purely scale-free one. Remarkably,
in fact, the average nearest-neighbor degree grows monotonically with $\lambda^{-1}$. One can
conclude that the more the distribution resembles a pure power-law, the less well connected are
the nodes, which might explain why $q$-exponentials are so ubiquitous.

\begin{figure}[t]
\includegraphics*[width=\columnwidth]{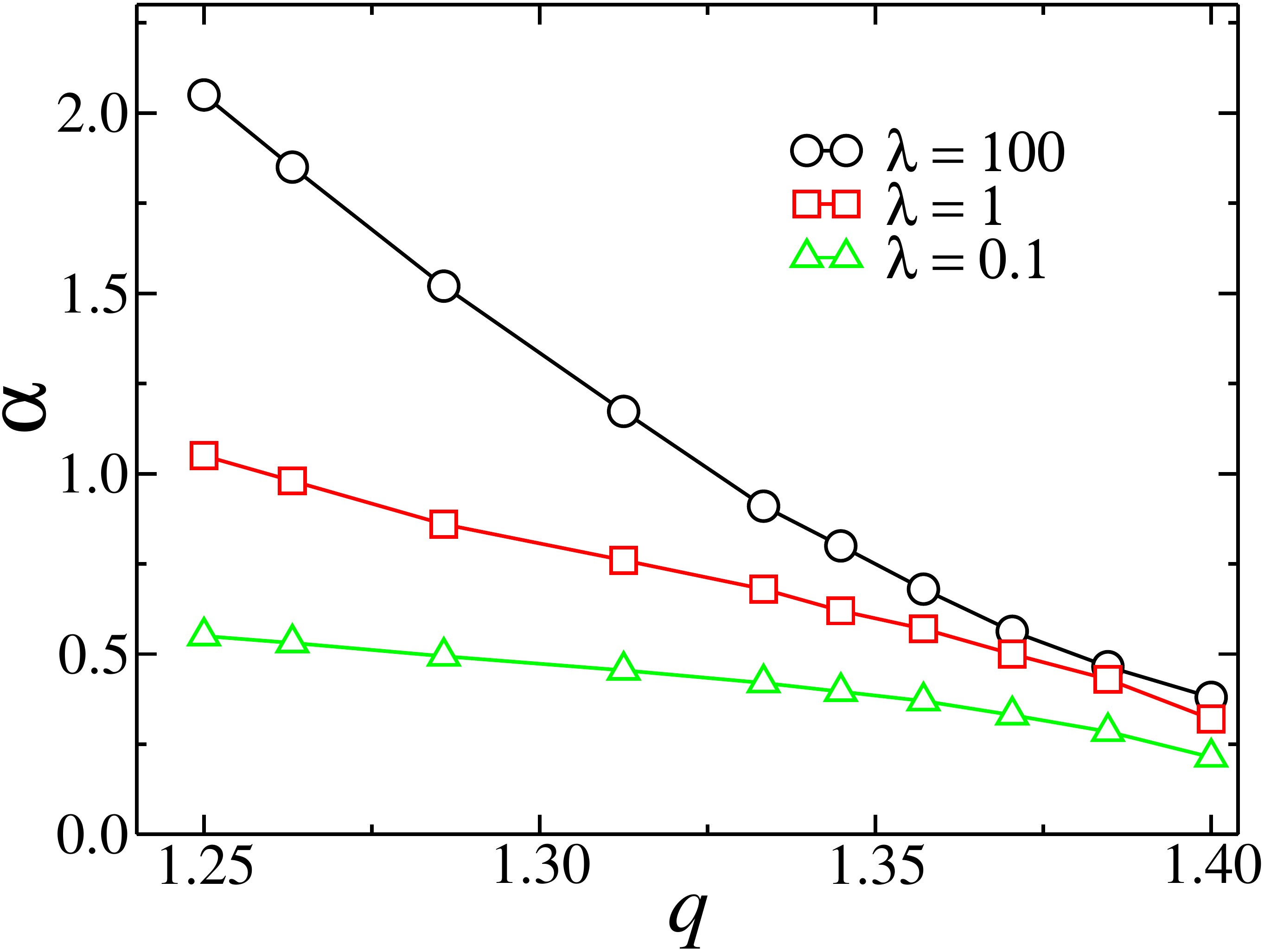}
\caption{The scaling exponent $\alpha$ as a function of $q$, for $\lambda=100$ (black circles),
$\lambda=1$ (red squares), and $\lambda=0.1$ (green triangles). Each  point is obtained
from a least-squares fit to the data, $\left\langle \ell \right\rangle \sim \alpha\log N$, with error bars being smaller than symbols.  The continuous lines represent guides to
the eye.}
\label{fig04}
\end{figure}

Random $q$-exponential networks also exhibit small-world behavior, which means that their
average shortest path increases logarithmically with network size,
$\langle\ell\rangle\sim \alpha\log N$, as shown in Fig.~\ref{fig03}.
Clearly the shortest path becomes considerably shorter as the degree distributions systematically 
deviate from the pure power-law, namely, for sufficiently large values of $\lambda$. Figure~4 shows the variation of the pre-factor $\alpha$ for different values of $q\equiv (1+\gamma)/\gamma$ in the range $1.25\leq q \leq 1.4,$ and $\lambda = 0.1,1,$ and $100$. For all practical purposes, this range of $q$, which corresponds to $2.5 \leq \gamma \leq 4.0$, covers all interesting power-law decays. As already mentioned, the case $\lambda=100$ is equivalent to a scale-free network with $P(k)\sim k^{-\gamma}$ and $k_{min}=2$.

\begin{figure}[t]
\includegraphics*[width=\columnwidth]{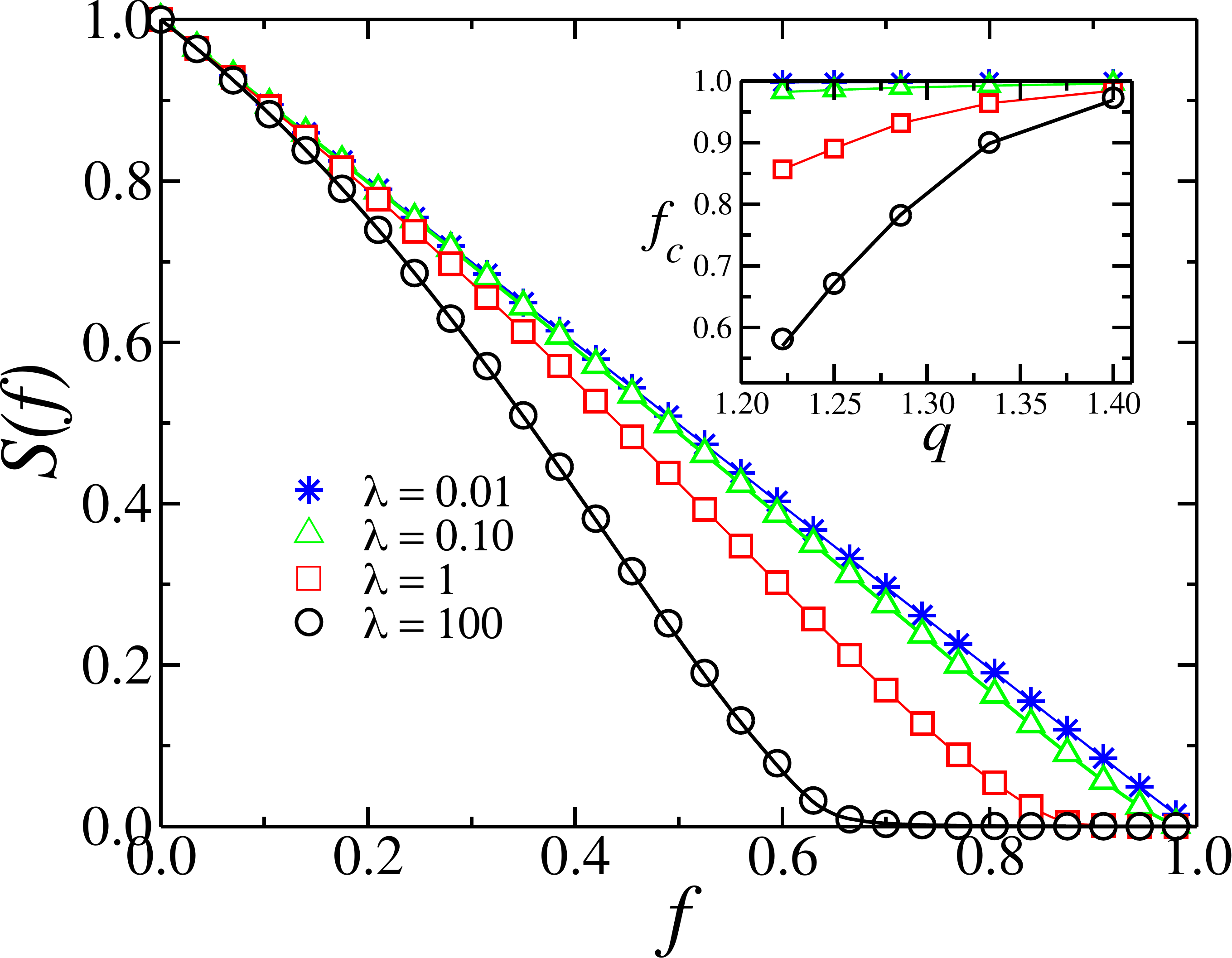}
\caption{The number of nodes in the largest cluster $S(f)$ as a function of the fraction $f$ of removed
nodes for random failures in $q$-exponential networks with $q = 1.25$, $\lambda=0.01$ (blue stars),
$\lambda=0.1$ (green triangles), $\lambda=1$ (red squares), and $\lambda=100$ (black circles). The results are
obtained for networks of size $N=320000$ by averaging over $200$ samples. The continuous lines for $\lambda = 0.01,0.1,$ and $\lambda=1$
represent guides to the eye, while for $\lambda=100$ the black solid line corresponds to  the result for random attacks on a pure power-law network. The inset shows the corresponding values of the fraction for which the largest cluster becomes negligibly small, namely, the critical fraction $f_c$ for random failures as a function of $q$.}
\label{fig05}
\end{figure}

A particularly important property for practical purposes is the robustness of networks against
random failures. In Fig.~\ref{fig05} we plot the number of nodes in the largest cluster $S(f)$ as
a function of the fraction $f$ of randomly removed nodes, for $N=320000$,  for $q = 1.25$, and various values
of $\lambda$. Our results show that, for this specific value of $q$, sufficiently large networks generated with
$\lambda \leq 0.1$ are completely robust, in the sense that they always exhibit a finite fraction of nodes $S(f)$ in
the largest cluster when subjected to a random attack, namely, $f_{c}=1$. On the other hand, for $q$-exponential networks
with $\lambda >1$, there exists a threshold $f_{c} <1$ above which the structure is completely disrupted, $S(f)=0$ for any value of $f \geq f_{c}$. 
In the inset of Fig.~\ref{fig05} we see how $f_c$ depends on
$q$ for different values of $\lambda$. Clearly, the closer a $q$-exponential network resembles a purely scale-free one, namely, for large values of $\lambda$, the more it is fragile against random failures. Furthermore, this effect is
dramatically amplified with the increase of the parameter $q$.

\begin{figure}[t]
\includegraphics*[width=\columnwidth]{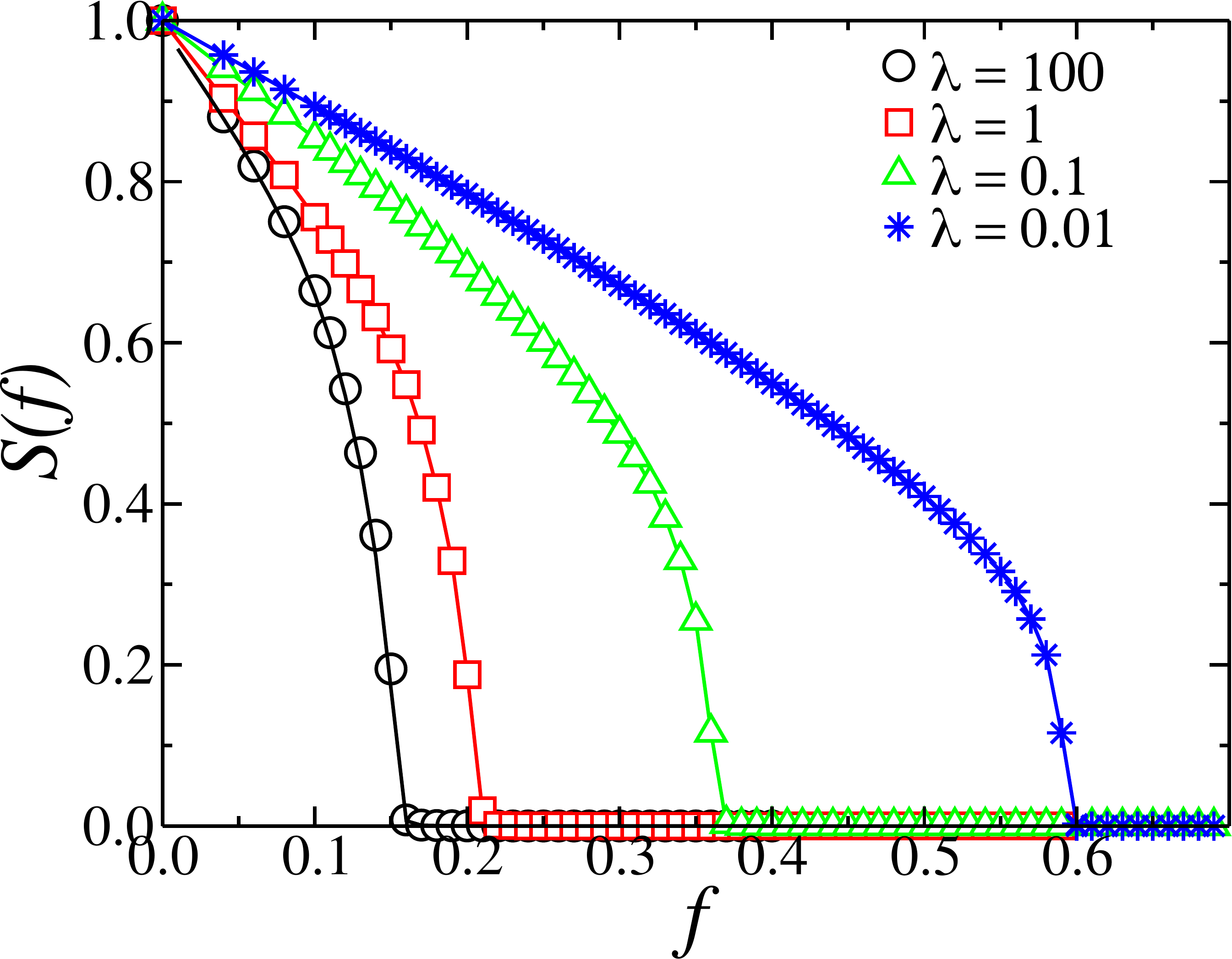}
\caption{The number of nodes in the largest cluster $S(f)$ as function of the fraction $f$ of removed
nodes for malicious attacks for $q = 1.4$, for $\lambda=0.01$ (blue stars), $\lambda=0.1$ (green triangles), $\lambda=1$ (red squares), and $\lambda=100$ (black circles). The continuous lines for $\lambda = 0.01,0.1,$ and $\lambda=1$ represent guides to the eye, while for $\lambda=100$ the black solid line corresponds to the result for malicious attack on a pure power-law network. The results are obtained for networks of size $N=320000$ by averaging over $200$ samples.}
\label{fig06}
\end{figure}

In Fig.~\ref{fig06} we show the variation of the size of the largest cluster $S$ as a function of the fraction $f$ of removed nodes, in the case of malicious attack for $q=1.4$ ($\gamma=2.5$) and different values of $\lambda$. As depicted, the $q$-exponential networks become less and less resilient as the crossover $\lambda$ increases, since a larger fraction of nodes needs to be removed before the critical point is achieved. This behavior persists up to a point at which the value of $\lambda$ is sufficiently large, so that the scale-free behavior of the degree distribution dominates. As a result, the $q$-exponential curve $S$ versus $f$ for $\lambda=100$ and the corresponding one generated from networks with pure scale-free distribution are perfectly coincident.

\begin{figure}[t]
\includegraphics*[width=\columnwidth]{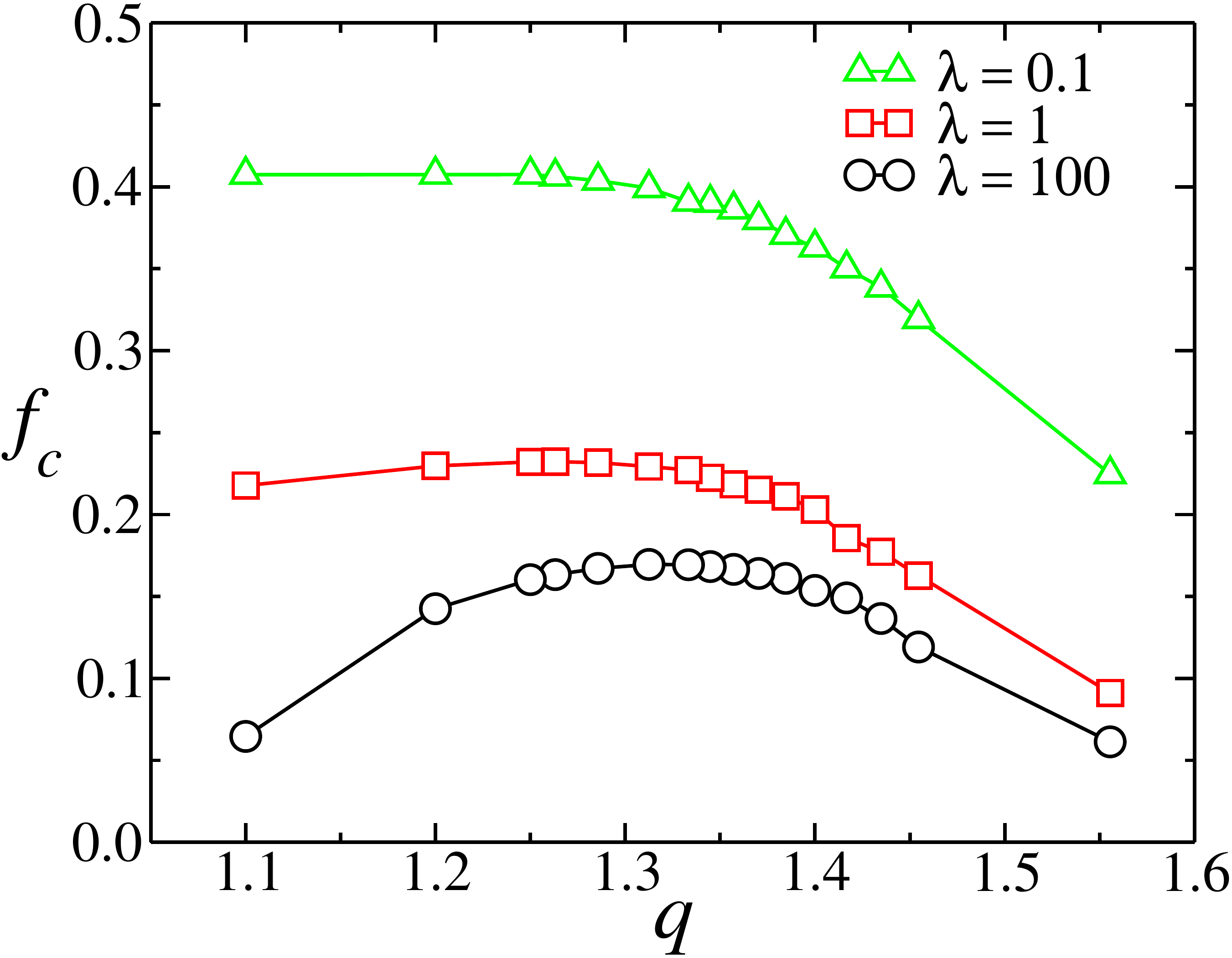}
\caption{The critical fraction $f_c$ for malicious attacks as a function of $q$, 
for $\lambda=0.1$ (green triangles), $\lambda=1$ (red squares), and $\lambda=100$ (black circles). 
The results are obtained for networks of size $N = 500000$ by averaging over $2000$ samples.
The fraction from which the largest cluster does not obey Molloy-Reed's criterion is the critical fraction $f_{c}$.
The $q$-exponential networks with smaller values of $\lambda$ are clearly more robust,
as a larger fraction of nodes needs to be removed to attain the critical point. The
continuous lines represent guides to the eye.}
\label{fig07}
\end{figure}

Figure~\ref{fig07} shows the dependence of the critical fraction $f_{c}$ on the parameter $q$, as determined by Molloy-Reed’s criterion~\cite{Molloy-Reed} and for different values of $q$. It is interesting to note that the behavior of $f_{c}$ changes substantially with $\lambda$. For instance, considering $\lambda = 0.1$ we see a plateau up to $q \approx 4/3$, followed by a decay. On the other hand, for $\lambda = 100$ we observe a clear a maximum in $f_{c}$ at $q \approx 4/3$. As mentioned, for $\lambda = 100$ the degree distribution approaches the form of a power-law, and this maximum in the critical condition is consistent with the expected for one purely scale-free networks~\cite{barabasiBook}. We note that this maximum is due to a compromise between two effects. For $q>4/3$ the degree distribution decays asymptotically as a power-law with controlling exponent $\gamma <3$, reaching $\gamma=2$ as $q$ approaches $3/2$.  At this limit removing just a few hubs results in a total breakdown of the network. At the other limit, as $\gamma$ diverges when $q \to 1$, the degree distribution is no longer heavy-tailed, decreasing rapidly. Moreover, if we set $k_{min}=2$, the generated network is already near the critical state. This behavior can be suppressed by imposing $k_{min}\geq 3$ in the case of pure scale-free networks~\cite{barabasiBook}, or by using small values of $\lambda$ in $q$-exponential networks.

\begin{figure}[t]
\includegraphics*[width=\columnwidth]{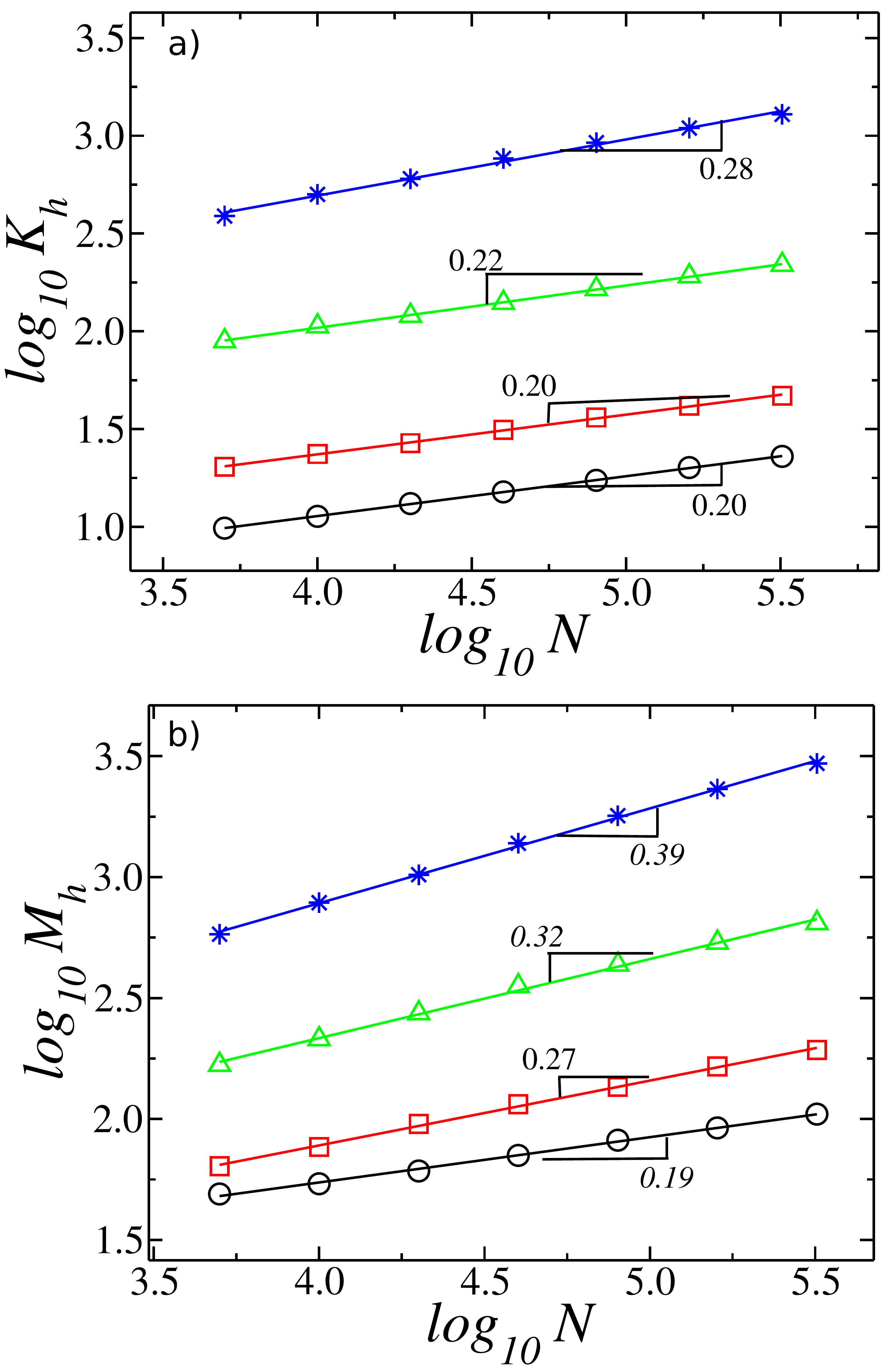}
\caption{(a) Highest $k_s$-core and (b) mass of the highest $k_s$-core versus $N$ for $q = 1.4$
and $\lambda=0.01$ (blue stars), $\lambda=0.1$ (green triangles), $\lambda=1$ (red squares), and $\lambda=100$ (black circles). The results are obtained by averaging over $10^{5}, 10^{4},10^{4},10^{4},10^{4}$, $10^{3}$, and $10^{3}$ samples of size $N=5000, 10000, 20000, 40000, 80000, 160000,$ and $320000$, respectively.}
\label{fig08}
\end{figure}

Next we extend the analysis of the  topology of $q$-exponential networks by investigating their hierarchical structure in terms of the $k_{s}$-core decomposition method~\cite{alvarez2006large,kitsak2010identification,pei2014searching,morone2019k,burleson2020k,Dorogovtsev,serafino2022digital}. The $k_{s}$-core of a graph $G$ is the largest connected subgraph in which all its nodes have a degree larger than or equal to $k_{s}$. To obtain the $k$-core, we remove all nodes with a degree less than $k$. Next, we scan if
some nodes still have a current degree less than $k$ and remove them. We repeat this check until no additional removal is possible. From this decomposition, we can define for each node a rank in the network, such that a node will be the more peripheral the smaller is its $k_{s}$. $k_{s}$-cores subgraphs are resilient against failure, since they preserve their convexity after $(k-1)$ random rewirings of edges or nodes. This kind of robustness tends to increase for the innermost nodes. Here we analyse the finite-size dependence of the highest $k_{s}$-core and its mass for different values of the parameter $\lambda$. From Ref.~\cite{Dorogovtsev}, we expect to obtain finite-size scaling as
\begin{equation}
k_{h} \sim N^{\delta},
\end{equation}
for the highest $k_{s}$-core, and 
\begin{equation}
M_{h} \sim N^{\Delta},
\end{equation}
for the mass of the highest $k_{s}$-core. We find that the exponents $\delta$ and $\Delta$ are functions of the distribution parameters $q$ and $\lambda$. Figure~\ref{fig08}a shows the finite-size dependence on the network size $N$ of the highest $k_{s}$-core in a double-logarithmic plot for $q=1.4$ and different values of $\lambda$. While the exponent
$\delta$ does not seem to depend noticeably on $\lambda$, the prefactor of the relation of Eq.(2) exhibits a considerable increase with $\lambda^{-1}$. As shown in Fig.~\ref{fig08}b, the 
dependence of the  mass of the highest $k_{s}$-core on the system size indicates that both the exponent $\Delta$ and the prefactor of Eq.(3) decrease substantially with $\lambda$. 
This shows that networks with a larger plateau (smaller $\lambda$) have a highest $k_{s}$-core that is bigger and has a larger $k_{s}$. This is in accordance with our previous findings 
that $q$-exponential networks become more robust the smaller $\lambda$.

In conclusion, we have for the first time generated unbiased complex networks
exhibiting $q$-exponential degree distributions with arbitrary parameter values.
These networks have several practical advantages with respect to pure scale-free
networks, namely, on one hand their nodes tend to be more homogeneously connected
and on the other hand they are more robust against failure, notably malicious attacks. 
These advantages might explain why $q$-exponential networks are so ubiquitous and found
in so many situations. In the future it would be interesting to also study dynamical properties of these networks,
like synchronization, epidemic spreading and opinion formation.

\begin{acknowledgments}
We  thank  Constantino Tsallis for helpful discussions and the  Brazilian  agencies  CNPq,  CAPES,  FUNCAP and the  National Institute  of  Science  and  Technology  for  Complex  Systems (INCT-SC) for financial support.
\end{acknowledgments}

\end{document}